# On reconciling quantum mechanics and local realism

Donald A. Graft
donald.graft@cantab.net

**ABSTRACT**

A necessary and natural change in our application of quantum mechanics to separated systems is shown to reconcile quantum mechanics and local realism. An analysis of separation and localization justifies the proposed change in application of quantum mechanics. An important EPRB experiment is reconsidered and it is seen that when it is correctly interpreted it supports local realism. This reconciliation of quantum mechanics with local realism allows the axiom sets of quantum mechanics, probability, and special relativity to be joined in a consistent global axiom set for physics.

## 1. INTRODUCTION

The apparent nonlocality of quantum correlation (entanglement) has confounded all attempts to reconcile it with other basic laws, such as Lorentz invariance. It is not hard to see why. The accepted quantum mechanics (QM) prediction for the probability of measuring both photons as 'up', for separated measurements of the correlated spin-1 singlet state at stations A and B, is believed to be given by a joint distribution, which we call $QM(AB)_J$, where J stands for joint and $\theta$ is the angle between the analyzer settings at A and B, $\alpha$ and $\beta$, respectively:

$$QM(AB)_J = P(AB|\alpha,\beta) = \tfrac{1}{2}\cos^2(\theta)$$

The A and B events are correlated as determined by the singlet state. The measurements taken in an EPRB experiment, however, are separated, so the measurement at A proceeds in ignorance of the analyzer setting at B and the outcomes at B, and vice versa. Therefore, we expect to measure the marginal probabilities at A and B, yielding:

$$QM(AB)_M = P(A|\alpha)P(B|\beta)$$

The expression $QM(AB)_M$ represents the predicted probability using the marginals $P(A|\alpha)$ and $P(B|\beta)$. Strangely, quantum physicists generally believe that the prediction $QM(AB)_J$ can be measured via separated sampling of A and B. But this leads to a contradiction, because it means:

$$QM(AB)_J = QM(AB)_M$$

Or equivalently:

$$P(AB|\alpha,\beta) = P(A|\alpha)P(B|\beta)$$

This is the condition for independence of A and B, but we know their dependence (correlation) was assured by the singlet state. We reach a *reductio ad absurdum* of the idea that a joint probability can be sampled by means of marginal (separated) measurements. Similar considerations apply for the anticorrelated singlet state.

The *reductio* can be avoided by supposing that when a photon is first measured at A, the companion photon at B is 'projected' to a state that yields a measurement compatible with $QM(AB)_J$. The prediction is in the form $P(A|\alpha)P'(B|\beta)$, where $P'(B|\beta)$ represents statistics from the measurement of projected photons. The *reductio* is dissolved, but at a terrible price: we have to accept superluminal effects of a strange kind for which no physical mechanism is known. We also have to ask which photon projects the other. Is it the first one to be measured? That is impossible to decide without a preferred reference frame when events have different orders in different reference frames. Even from an engineering perspective, it is challenging to conceive a protocol ensuring that one and only one of the photons projects the other, and the resulting mechanism has to be regarded as fanciful.

We see then that the axioms of QM (when interpreted in the conventional way that applies the joint probability prediction to separated measurement situations) conflict with those of special relativity (SR) because QM nonlocal influences are not Lorentz invariant and they require a preferred reference frame, and/or those of probability (P) because QM implies that a joint probability can be measured via the marginals. We cannot try to live with these conflicts because the resulting combined axiom set that we use to describe nature is not self-consistent, and when axioms are inconsistent contradictions are easily generated and we have no way to distinguish sense from nonsense. We assume the axioms of QM alone are consistent, or a consistent set can be defined, although it has been sporadically argued that QM itself is inconsistent.

Nonlocality appears to be the only aspect of QM preventing the consistent combining of the axiom sets of QM, SR, and P. Schrödinger[1] famously stated "I would not call [nonlocal entanglement] *one* but rather *the* characteristic trait of quantum mechanics, the one that enforces its entire departure from classical lines of thought." Feyman[2] discusses simulation of quantum mechanics by a classical computer but stumbles when addressing nonlocal correlations: "That's all. That's the difficulty. That's why quantum mechanics can't seem to be imitable by a local classical computer." Spreeuw[3] demonstrated classical analogs of local entanglement but reported he was unable to demonstrate a classical analog of nonlocal entanglement. Finally, Orlov[4] was able to demonstrate many aspects of quantum computation using classical building blocks but failed to imitate nonlocal quantum effects using his classical blocks.

It has been argued that negative Wigner functions (or other metrics) associated with entanglement have no classical counterpart but negativity is associated with contextuality rather than entanglement[5]. Classically contextual systems are abundant, so there is no fundamental conflict due to negativity. Quantum nonlocalists may also argue that contextuality requires a new quantum probability theory because a sample space is said not to exist for contextual systems, and that standard Kolmogorovian probability therefore cannot be applied. However, any contextual system will be represented by a set of sample spaces, one per context. That is not the same as not having a sample space! Contextuality is not a problem for standard probability theory, and contextuality can be treated identically in the classical and quantum representations.

We seek, therefore, a way to eliminate nonlocality from QM, and we do that by simply accepting that a joint distribution cannot be sampled by means of separated (marginal) measurements. One must use $QM(AB)_M$ instead of $QM(AB)_J$ for predicting the measured correlation. This prediction can be made via partial traces or reduced density matrices in a manner completely analogous to that of marginalization and conditional probabilities in standard probability theory. If this is accepted, then a very small reinterpretation of QM can reconcile QM and local realism: using the marginals versus the joint probability in separated measurement situations (exactly as in classical probability). Specification of what are separated measurements is a delicate matter but has a satisfactory answer developed in this paper.

Of course the experiments are obstacles for advancing the reconciliation program. Modern consensus is that the results of EPRB and other experiments confirm that the $QM(AB)_J$ prediction is clearly obtained. Nevertheless, it is argued that the experiments have been misinterpreted. An analysis of a careful and oft-cited EPRB experiment by Weihs *et al*[6], hereafter referred to simply as Weihs, reveals a *plausible* local realistic account of the experiment's results. The results are shown to approximate the $QM(AB)_J$ solution just as the Weihs experiment does, but as a result of local realistic mechanisms rather than the sampling of a joint probability via its marginals implied by the $QM(AB)_J$ prediction. While attempting to deconstruct all of the experiments can quickly turn into an exasperating game of *whack-a-mole*, we will see that the mechanism in effect in the Weihs experiment has broad applicability, so a plausible local realistic account of the Weihs experiment goes a long way toward clearing the path for reconciliation of QM and local reality.

In consequence of the foregoing, this paper's goal is twofold, first, to characterize separated measurements and show that the joint probability formula cannot be applied to separated measurements, and second, to show that the experiments have been misinterpreted and that nonlocal entanglement is an error. As a result, locality is restored and our global axiom set for physics becomes fully consistent.

## 2. SEPARATED MEASUREMENTS

It is easy to devise a local system that embodies the sampling of a joint PDF and generates quantum statistics. Consider first a simple physical system that embodies a preparation followed by two dichotomic measurement results A and B as shown in Figure 1. A disk of unit radius is prepared by partitioning it into four sectors according to a parameter $\theta$. Each resulting sector is labeled with outcomes for A and B as shown in Figure 1. We may think of spinning an arrow attached to the center of the disk and noting the location on the circumference that the arrow points to when it comes to rest. Effectively we have generated a random number $\lambda$ in the range $0-2\pi$ and used it to determine the location. The outcomes for A and B are then read off from the sector containing the arrow position as shown in the figure. When the measurements are repeated, a sequence of outcomes is obtained for A and B and we can correlate them in the usual way. For the preparation shown in Figure 1, we obtain statistics that exactly reproduce the quantum correlations of the anticorrelated singlet state. The physical system clearly embodies the joint PDF for the A and B outcomes, and the quantum statistics are obtained simply by sampling the joint PDF. This simple model should serve as a useful tonic for those who mistakenly believe that correlations of dichotomic functions cannot be harmonic, and thus cannot conform to the predictions of QM.

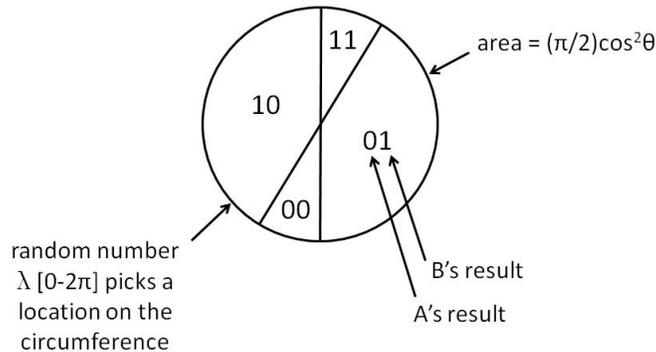

Figure 1. Measurement results for A and B are produced by the sampling of a joint PDF embodied as a physical disk. The preparation can be considered to be the construction and setup of the disk.

The physical system can be separated in space and time without destroying the ability to successfully sample the joint PDF. Consider a refinement of the system as shown in Figure 2. The original disk is split into two, retaining the respective outcome labels for A and B. As long as shared randomness is used for the sampling (think of each disk being indexed by a single spinning arrow), the correct joint PDF defined in Figure 1 is still successfully sampled. The disks can be separated in space and the measurements can be made at different times, using the shared random variable, without affecting the results. We conclude counter intuitively that the naïve view that simple physical separation in space or time defines a separated measurement is incorrect, and we must look further for the essential features of separated measurements.

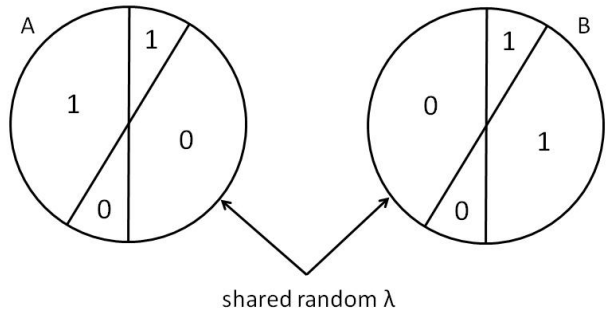

Figure 2. The original disk has been split into two physically separated disks. Shared randomness is used for sampling the disks. The measurements at each side may also occur separated in time. The joint PDF of Figure 1 is still successfully sampled even though the measurements at A and B are physically separated in space and time.

If shared randomness is not available to the two sides of the separated system, then the joint PDF cannot be sampled. Consider a further refinement of the system as shown in Figure 3. The measurements are now generated without shared randomness, i.e., different random variables are used at A and B (think of each disk having its own spinning arrow). The joint PDF now cannot be successfully sampled and the result is a function of the marginal probabilities. It is tempting to assume that when systems are physically separated in space they no longer have access to shared randomness, but this is not necessarily so. The source events themselves could be distributed in a random distribution and this randomness would be transmitted to both sides. Shared randomness from the source is not the only possibility; for example, suppose the systems are separated by several kilometers but they both observe a specific location on the sun and read off the intensity variation and treat that as a shared random variable. So we need to have a good understanding of the system and its physics before we can know if shared randomness is available and if it is indeed relevant to the sampling. For example, we might suppose that a light packet ('photon') actually is represented by one of the disks in Figure 3 spinning at a fixed rate (this conception brings to mind Feynman's idea of little stopwatches in his description of quantum electrodynamics[7]), and then the measurement is indexed by a fixed arrow at 0 degrees. Now, if as is almost certainly the case, we take our measurements at independent times at A and B, i.e., the measurement times are not synchronized, then we cannot successfully sample the joint PDF.

The idea of broken shared randomness cannot be definitively applied to the EPRB experiments, however, for at least two reasons. First, there is still uncertainty about the actual nature of light, so asserting such an interpretation of the experiments may rely upon assumptions not yet evident. Exhaustive identification of all the shared randomness in an experiment and its relevance to the detection processes is difficult. Second, it is doubtful that current light detectors are sensitive to the phase of the light, because the time scale of the energy integration period leading to a detection event is large compared to the wavelength of the light. One could *try* to build a case that the experiments involve separated measurements due to lack of shared randomness, but at this time it would be speculative and inconclusive. Fortunately, we need not assert that this mechanism is in play, although we certainly don't exclude the possibility, because a second mechanism that leads to separated measurements that we can unquestionably assert is in play.

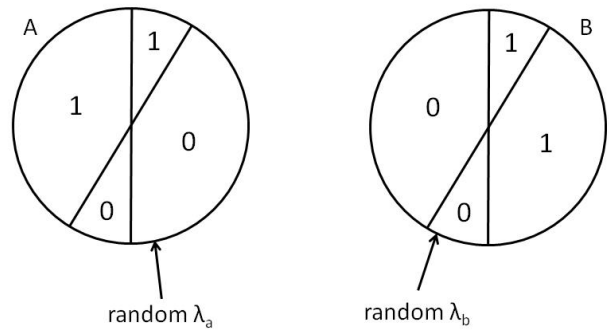

Figure 3. The measurements are generated without shared randomness, different random variables are used at A and B. The joint PDF can no longer be successfully sampled.

When parameters affecting the measurements are available to both sides, the joint PDF can be successfully sampled. Consider a further refinement of the system as shown in Figure 4. The use of shared randomness has been restored because we cannot definitively show that shared randomness is not available to the two sides in the EPRB experiments. The system is now the same as the one in Figure 2, but with the single parameter $\theta$ replaced by the difference of two parameters $\alpha$ and $\beta$. The two disks are identical so each disk must have been prepared with knowledge of both $\alpha$ and $\beta$. For this arrangement the joint PDF is still successfully sampled, and so, like the system shown in Figure 2, we cannot call this a separated measurement situation.

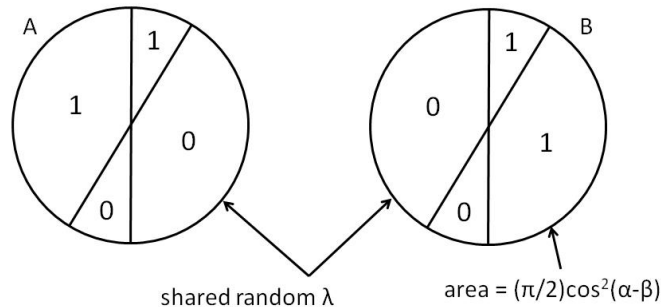

Figure 4. The parameter $\theta$ in Figure 1 is replaced by the difference between two parameters $\alpha$ and $\beta$. The joint PDF can still be successfully sampled.

Now the fun begins. When parameters affecting the measurements are not available to both sides, the joint PDF cannot be successfully sampled. We suppose that A does not know about $\beta$, and B does not know about $\alpha$, as depicted in Figure 5. It is clear that the original joint PDF of Figure 1 cannot be sampled. Of course each side could be very lucky and guess the other side's parameter every time, but logically that is the same as each side knowing the other side's parameter. If we were tasked with predicting the correlations for such an experiment, we would have several possibilities for treating the unknown parameters. We could for example ignore the unknown parameters completely (equivalent to assuming they are 0), as shown in Figure 5. We could also assume any fixed values, we could assume random values, or we could integrate over the possible values. But none of these options leads to a proper sampling of the original joint PDF. The parameters $\alpha$ and $\beta$ of course correspond to the measurement angles at the two sides of the EPRB experiments, and so we must conclude that the EPRB experiments involve separated measurements. The quantum joint formula is customarily stated as $\frac{1}{2}\cos^2\theta$ rather than $\frac{1}{2}\cos^2(\alpha-\beta)$, which amounts to surreptitiously sharing angle parameters, and so one might understand how an uncritical perusal of the function ½cos²θ may not expose the definitive measurement separation inherent in this situation.

There is a special case of interest that successfully samples the original joint PDF, but which *appears* not to share parameters between the sides. If side B chooses β to be 0, and side A assigns its area as πcos²α [was previously (π/2)cos²α], then the original joint PDF is successfully sampled for any value of parameter α. However, this removes freedom of choice of the measurement angle from side B, or it involves a surreptitious communication of a redefinition of the origin of the angular frame of reference (to define B's chosen parameter as 0) equivalent to sharing of parameters.

The account of measurement separation here involves only considerations in probability, not any localization in time and space. Time and space enter only through their constraints on the availability of required shared randomness and parameters. For example, suppose side A does her measurement and shouts to side B revealing her measurement angle. As long as side B is within hearing distance (and they have some shared randomness – side A can shout that too if necessary), side B can easily generate results yielding quantum correlation. If side A and side B move too far apart, side B can no longer hear the shouts and so cannot generate quantum results. Space has constrained the possibilities for sharing. A similar effect can occur in time. With the addition of space and time in this manner, we arrive at a satisfying and useful way to define and distinguish separation and localization, and to see how they interact.

The considerations thus far adduced allow us to characterize and identify separated measurement situations and to recognize that the EPRB experiments involve such separated measurements. That being so, the standard quantum joint prediction $QM(AB)_J$ cannot be applied and only the prediction using the marginals $QM(AB)_M$ can be applied. Stated simply, the standard quantum prediction for correlation of separated measurements is incorrect and it leads to all of the difficulties previously recounted. As noted earlier, however, the experiments appear to show that the joint prediction results are obtained. We turn therefore to an analysis of an important exemplar of the EPRB experiments[6], and show that the experiments have been misinterpreted, that in fact the joint prediction is not obtained, and that nonlocal entanglement is an error. Framing the debate as quantum mechanics versus local realism is a false, misleading, even inflammatory apposition; the true debate is over whether the joint prediction can be applied to separated measurements or not. As we have seen, basic classical probability theory tells us that it cannot. Making this small but needed adjustment for separated measurement leaves the essence of quantum mechanics intact.

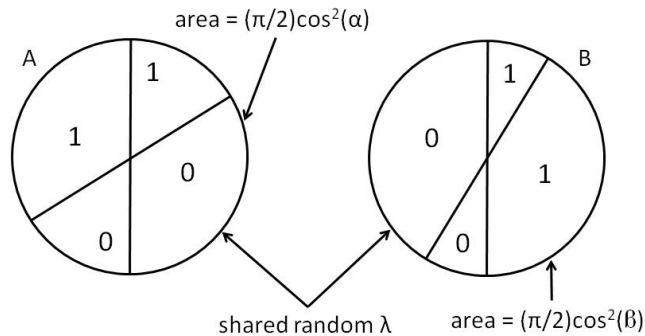

Figure 5. When A doesn't know β and B doesn't know α, the original joint PDF cannot be successfully sampled.

## 3. THE EXPERIMENTS

In an earlier paper by this author[8] a local realistic account of the Weihs EPRB experiment[6] was described. To limit this conference paper to a reasonable length and to new material, readers are referred to the cited paper for a full description of the local realist account and its correspondence to the important and pioneering Weihs experiment. We focus here only on the important concepts and the results of simulations of the local realist account. Readers should also refer to the references contained in the cited paper to properly appreciate the important pioneers in this field. The author here acknowledges particular indebtedness to (in addition, of course, to the classical giants Schrödinger and Einstein) Weihs *et al*, Caser, Marshall, Santos, Fine, Larrson, Hofer, Khrennikov, Adenier, and De Raedt *et al*. The latter two are presenting at this conference and we can be sure that they will advance our understanding in important ways, as they have done many times in the past.

The primary reports on the results of EPRB experiments all neglect to consider, report, and account for the choices for calibration of important parameters of the light detection apparatus, specifically and importantly, the detection thresholds used during analysis and processing of the analog detection traces to produce the dichotomic measurement outcomes. In the cited local realist account[8], when the thresholds are properly and independently calibrated, only classical correlations are observed. When one of the thresholds is miscalibrated, a full range of results spanning classical to super-quantum correlations (with full rotational invariance) can be obtained, depending on the extent of the miscalibration. The mechanism for this effect is straightforward unfair sampling of a deterministic device subject to Malus's Law as shown in the cited paper.

One of the earliest careful EPRB experiments by Holt and Pipkin[9] failed to show quantum correlations and therefore weighed strongly in favor of the local realist account. Instead of some unidentified systematic error being responsible for their results, as is usually suggested, it is easier to believe that Holt and Pipkin *properly* calibrated their apparatus. Holt and Pipkin were subjected to strong scientific peer pressure, and they never formally published their results.

According to this view, the subsequent experiments were inadvertently miscalibrated. The experimenters sought to calibrate their experiments with the aim of verifying hypothesized nonlocality, and they mistakenly believed that the choice of threshold parameters is uncritical (as long as the chosen thresholds exclude the background noise). The claims here are bold, so readers are again referred to the cited paper for support. All the experiments that appeal to detection thresholds are potentially subject to the mechanism described.

The recent Giustina *et al* experiment[10], which claims to fully exclude all unfair sampling, opens an important and interesting new line of defense against the detection loophole. But unfair sampling is not needed to violate Giustina *et al*'s single-channel variant of the Eberhard inequality. The Eberhard derivation relies upon two crucial assumptions: the first obeys the law of large numbers and becomes statistically true for a large number of events, and the second is false (in real experiments) and is not subject to the law of large numbers. Indeed, a following paper will demonstrate violation of the Giustina *et al* inequality consistent with the experiment using a simple semiclassical model.

Let us recall the results of the computer simulation of the local realist account[8]. Before proceeding let us realize that although the account uses dual-channel detectors at each side, as does the Weihs experiment, the described effect also applies to single-channel detectors. For any given source event in the dual-channel case, we see at a given side either a miss by both detectors, a double hit by both detectors, or a single hit at one detector. We cannot have doubles in the single-channel case, but we can still have misses. Significantly, the account here of the Weihs data relies on misses.

When both sides are correctly calibrated classical results are obtained. Figure 6 shows the match probability curves resulting from the local realist simulation when the detection thresholds at both sides are each correctly calibrated to one half the light pulse energy (0.5 in a normalized scale). To generate the curves, the measurement angle at side A is set to 0 and the measurement angle at side B is scanned over the range 0-$\pi$. The threshold does two things. It excludes low-level noise when it is set higher than the noise level. But just being above the noise is not sufficient. The threshold must ensure fair sampling. Only a threshold of 0.5 both excludes the noise and performs fair sampling by not discarding significant detection events in a pattern governed by Malus's Law. Readers have undoubtedly noticed that the results in Figure 6 are classical and various inequalities are not violated.

When one side is grossly miscalibrated, super-quantum results are obtained. Figure 7 shows the match probability curves resulting from the local realist simulation when the detection threshold at side A is correctly calibrated to 0.5, while side B is miscalibrated to 0.92 (normalized to the signal energy). The system is still fully rotationally invariant. As long as one side is correctly calibrated, the system delivers full rotational invariance. If both are miscalibrated, then rotational invariance is destroyed and modulation of the number of total coincidences is seen as B is scanned over 0-$\pi$. One can thus go astray when trying to draw inferences from the presence or absence of rotational variance if one lacks an understanding of the effects of the detection thresholds on the rotational symmetry.

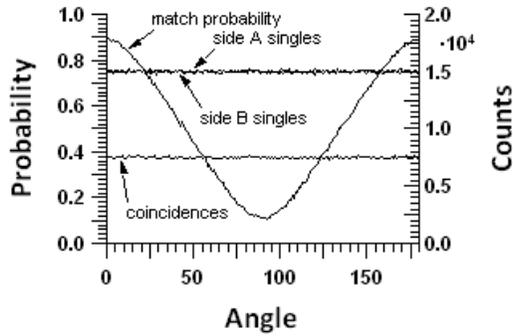

Figure 6. Classical results of the local realist simulation when both sides are correctly calibrated for a threshold of 0.5 (normalized to the signal energy).

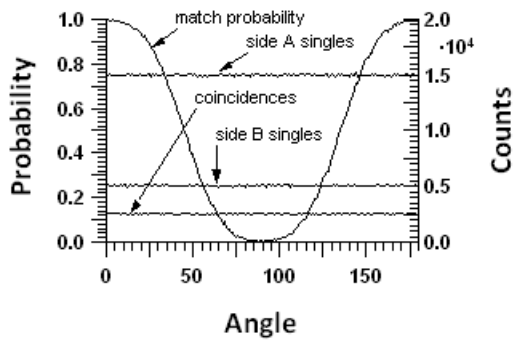

Figure 7. Super-quantum results of the local realist simulation when side A is correctly calibrated to 0.5, while side B is *miscalibrated* to 0.92 (normalized to the signal energy).

Between the two extremes of classical and super-quantum calibration, we can easily find a threshold value of 0.75 for side B (with side A left correctly calibrated) that produces the quantum results shown in Figure 8. The curves have high visibility and there is full rotational invariance. The glaring give-away to miscalibration is the large and predictable difference between singles counts between sides A and B seen in Figure 8. Notice that in Figure 6, where both sides are correctly calibrated, the singles counts are approximately equal at the two sides. The point to notice is that the greater the extent of the miscalibration, the greater the difference between singles counts at the two sides. In the super-quantum calibration (Figure 7) there is a very large counts asymmetry. This artifact is clearly seen in the Weihs experimental data[6]. Adenier and Khrennikov[11] present an analysis of the Weihs data in which the observed counts asymmetry is seen to be close to that observed in the local realist model when it is calibrated for 0.5/0.75 as described (Figure 8).

Although the EPRB experiments are incomplete due to their failure to report on different calibration domains, they remain capable in principle of testing whether the joint probability formula applies or not. Here we make no reference to testing quantum mechanics versus local realism, a false apposition, nor to deciding whether quantum mechanics is complete, a red herring. As long as the detectors are all properly and independently calibrated, the experiment can be decisive. Even low detection efficiency (5% in the Weihs experiment) is not a problem because nothing in quantum mechanics allows one to say that randomly discarding events can change the qualitative results. It is as if one simply performs a shorter experiment. So if classical results are seen with symmetric, correctly calibrated sides, then the joint probability formula *is not* correct. If fully rotationally invariant quantum results are seen, then the joint probability formula *is* correct. But we have shown earlier that the joint probability formula cannot be correct.

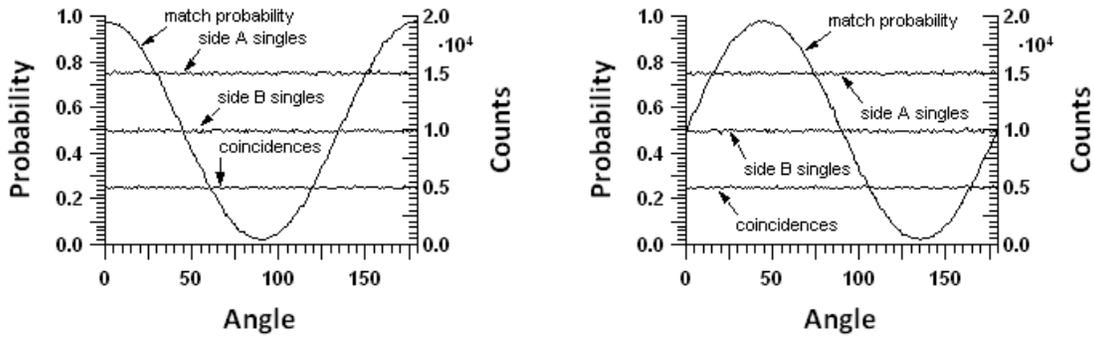

Figure 8. Quantum results of the local realist simulation with threshold 0.5 at A and 0.75 at B when *left*: side A's measurement angle is 0, while B's is scanned over 0-π, *right*: side A's measurement angle is π/4, while B's is scanned over 0-π.

## 4. DISCUSSION

Readers are directed to the extensive discussion contained in the previous paper[8]. Some speculations germane to a further interesting line of thought are reported here.

The crucial role of detector thresholds and proper apparatus calibration has been demonstrated, but the model presented here remains open to a charge that it is unphysical in an important respect. Local realist accounts of EPRB experiments, including this one, typically appeal to a random isotropic distribution of some source light property, such as phase of the electric field, polarization, etc. In the model here, the paired photon polarizations in EPRB are assumed to be emitted orthogonal to each other with the pair randomly oriented over 0-2π. In the EPRB experiments with parametric down conversion (PDC) light sources, however, the polarization of the source photon pairs is constrained to a single fixed H/V basis (in the classical interpretation and for a simplistic account of PDC with a linearly polarized pump laser), providing only 4-fold rotational symmetry, rather than the fully isotropic light source assumed in the local realist models. The local realist models therefore arguably fail for this PDC light, and indeed, simulations show pathology when side A's measurement angle is set to an odd multiple of π/4 (bisecting the H/V basis) while side B's angle is scanned. The Weihs experiment, however, displays clean 8-fold rotational symmetry.

Ironically, quantum mechanics cannot even formulate this objection, because the emitted photons are described by the singlet state, which is fully rotationally invariant and represents all that can be known or said about the photons. That doesn't get local realists off the hook, however, because a *plausible* local realist account must address this important line of analysis. Some progress has been made. A paper in progress by this author shows that the local realist account described here functions for reduced rotational symmetries. For example, the pump laser intensity can be calibrated such that the pair number statistics of the PDC light contain 1- and 2-pair events. 2-pair events contribute an additional component of rotational symmetry at π/4, and a computer simulation shows that 8-fold symmetry results. The Weihs *scanblue* data reports only results at 8-fold angles (for one side while the other is scanned). Importantly, the simulation with 8-fold symmetry shows the same dependence on the detector thresholds as the fully isotropic model, so our identification of the smoking gun in the Weihs experiment demonstrated for the isotropic model remains valid for models with reduced symmetries. The Weihs experiment is again incomplete due to its failure to report on the rotational spectrum of the source pairs and their light pulse (photon) number statistics.

The study of rotational symmetries created by different photon pair creation statistics and other sources of rotational symmetry is an important further direction for study of local realist models like the one presented here, and recognition of these mechanisms obliges us to perform proper tomography of the light source with sufficient rotational granularity to properly interpret the experiments. Another potentially interesting area to explore is the role of parasitic optical effects that might affect observable symmetries, such as flare (Figure 9), glare, internal reflections, etc. We know essentially nothing about any of these things in the Weihs experiment, and so the local realist account presented here simply marginalizes them away by assuming a fully isotropic light source distribution.

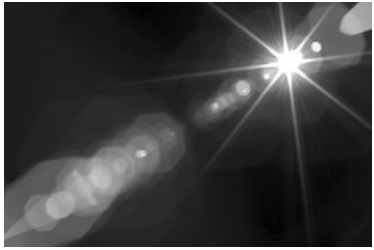

Figure 9. Typical lens flare showing 8-fold symmetry. This particular rotational modulation may be due to an adjustable diaphragm in the optical path. [Image from *Wikimedia Commons* free use collection.]

Finally, the argument of this paper is strengthened and complemented by recent important foundational work of De Raedt *et al.*[12] Because QM and classical local realism are not in conflict, as shown in this paper, the applicability of QM as an optimal form of inference can apply equally to classical situations, because those situations also require robust logical inference under uncertainty, as described by De Raedt *et al*. A person modeling or explaining a physical effect has the discretion to apply QM and/or classical representations in combination as needed. Due to the axiomatic consistency we have shown, one can place the Heisenberg cuts wherever they are needed. There are no paradoxes, mysteries, or difficulties there.

## 5. CONCLUSION

Five important themes have been manifested:

1. A plausible local account of the Weihs experiment is presented and demonstrated with a computer simulation.

2. Failure of rotational invariance is shown not to be a necessary outcome of unfair sampling. The Weihs experiment shows clear rotational invariance, and no plausible model had duplicated that. The model described here does so and identifies a 'smoking gun' in the experiment.

3. Device calibration and reporting of all apparatus parameters is shown to be critically important. That has clear implications for proper experimental design of EPRB-like experiments.

4. High detection efficiency is shown to be unnecessary to obtain definitive results from EPRB experiments. This is significant because the debate has thus far been distracted by a misguided focus on efficiency.

5. Most importantly, quantum mechanics is shown to be compatible with local realism, by means of correct handling of separated systems. We cannot use the joint probability formula for cases of separated measurements; instead we use the marginals (partial traces or reduced density matrices) together with whatever priors we have from an understanding of the system. Specification of what are separated measurements is delicate but has been adequately addressed here. If we accept this small reinterpretation of quantum mechanics, nonlocality is eliminated. The experiments when correctly interpreted confirm the local realist position. Nonlocal entanglement is seen to be an error. The rest of quantum mechanics remains intact, and remains highly valued as a powerful probability calculus for observables. Without nonlocality to contend with, we can recruit powerful classical ideas, such as semiclassical radiation theory, stochastic dynamics, classical noncommutativity/contextuality, measurement effects on state, etc., to augment or complement quantum mechanics. The modified quantum mechanics lives in peaceful harmony with the local realist conception.

## ACKNOWLEDGEMENTS

The author thanks Gregor Weihs for making available the raw data files from his experiment[6] and for valuable discussion of the ideas contained here; Hans De Raedt for encouragement and helpful discussion; Arthur Fine for helpful feedback and for stimulating me to think deeper about the Giustina *et al*[10] results; and finally SPIE for sponsoring this delightful conference that provides such an invaluable forum for discussing the foundations of quantum mechanics and the nature of light.